# Quantum Computer Controlled by Superconducting Digital Electronics at Millikelvin Temperature

Caleb Jordan[1]†, Jacob Bernhardt[1]†, Joseph Rahamim[2], Alex Kirichenko[1], Karthik Bharadwaj[2], Louis Fry-Bouriaux[2], Aaron Somoroff[1], Katie Porsch[2], Kan-Ting Tsai[1], Jason Walter[1], Adam Weis[1], Meng-Ju Yu[1], Mario Renzullo[1], Jerome Javelle[2], Chris Checkley[2], Oleg Mukhanov[1], Daniel Yohannes[1], Igor Vernik[1], Shu-Jen Han[1]*

[1]Seeqc, Inc.; Elmsford NY, 10523, USA.

[2]Seeqc UK; Unit Et.5.01 Cargo Works, 1-2 Hatfields, London, England, SE1 9PG

*Corresponding author. Email: sjhan@seeqc.com

† These authors contributed equally to this work

**Current superconducting quantum computing platforms face significant scaling challenges, as individual signal lines are required for control of each qubit. This wiring overhead is a result of the low level of integration between control electronics at room temperature and qubits operating at millikelvin temperatures, which raise serious doubts among technologists – can utility-scale quantum computers be built? A promising alternative is to utilize cryogenic, superconducting digital control electronics that coexist with qubits. Here, we report the first multi-qubit system integrating this technology. The system utilizes digital demultiplexing, breaking the linear scaling of control lines to number of qubits. We also demonstrate single-qubit fidelities above 99%, and up to 99.9%. This work is a critical step forward in realizing highly scalable chip-based quantum computers.**

Researchers and engineers developing superconducting quantum processor units (QPUs) have recently made significant progress toward the realization of quantum computers (QCs) that can potentially outperform classical computers in certain areas [1–3]. The recently demonstrated “below-threshold” error-corrected logical qubits, although performed on a system in the order of 100 physical qubits [4], provides a strong motivation for developing large-scale QC systems. On the path towards a fault-tolerant QC which consists of 100,000 to 1 million physical qubits, one key requirement is the extremely tight integration of qubits and control electronics to provide the necessary scalability. Current state-of-the-art QPUs rely on a brute-force scaling approach where the superconducting quantum chip operating at the millikelvin stage inside the dilution refrigerator (DR) is connected to the electronics at room temperature (RT) via many signal wires (Fig. 1A, left). This approach requires numerous racks of complex RT electronics for QPU operation, and has clear limitations – the finite cooling capacity of the DR to remove the heat generated from wires and analog components such as attenuators [5], the insufficient space inside the DR to accommodate these wires and components, and the tremendous footprint and energy consumption of RT electronics and large DRs. Therefore, researchers have explored more integrated solutions by inserting complementary metal-oxide-semiconductor (CMOS) based mixed-signal control electronics at cryogenic temperatures (Fig.1A, middle). Although the form factor can be significantly reduced in this cryo-CMOS implementation, the operation principle is largely the same as the conventional RT control scheme which relies on generating precisely shaped microwave pulses [6,7]. Ref 4 describes in detail such an implementation for the control of two qubits in a Google Sycamore quantum processor. To ensure that the gate error rate is below the quantum error correction threshold [4], high resolution (10 to 14-bit) digital-to-analog converters are required for XY and Z controllers to minimize the leakage and qubit frequency errors. The complexity of these control circuits (thus high transistor count), together with the large power consumption of CMOS technology result in ~2 to 23 mW/qubit [6,7] and require them to be thermalized at a higher temperature and connected to the quantum chip using ultra high-density superconducting cables, which still results in a significant I/O bottleneck.

One solution for scaling up solid-state QCs is to develop an active QPU where the control electronics and quantum layer co-exist at the same temperature, based on a chip-stacking architecture as shown in the right plot of Fig. 1A [8]. Communication between the two layers can be established via either galvanic contacts (superconducting micro-bumps) or via capacitive or inductive couplings through vacuum. Such an active QPU significantly reduces the aforementioned I/O overhead. Since control signals are generated locally, signal multiplexing and/or demultiplexing can be implemented to route these signals. However, to satisfy the ultra-low power requirement at the millikelvin stage, the lower energy efficiency of cryo-CMOS technology and large, complex circuits to implement conventional RF control techniques pose fundamental limitations. To circumvent these limitations, superconducting Single Flux Quantum (SFQ) technology with digital qubit control schemes that significantly reduce circuit complexity is an attractive alternative platform for building control electronics, due to its low energy dissipation and ultra-fast operation [8].

SFQ technology is based on Josephson-junctions and superconducting inductors, with classical bits represented by the presence or absence of single-flux-quanta in a loop [9]. The propagation of these flux-quanta through a circuit is accompanied by the 'switching' of a junction, which results in a short voltage pulse with an area quantized by $\int V(t)\mathrm{d}t = \Phi_0$. For charge-based qubits such as transmons[10], single-qubit control can be implemented by coupling these SFQ voltage pulses to the qubit through a small capacitance [11]. Each pulse delivers a fixed amount of energy to the qubit, rotating the qubit state about the Bloch sphere by an amount $\delta\theta$ (Fig. 1B, top). Primitive gates such as $X/2$ are realized by applying $n_p$ SFQ pulses at the qubit frequency or a subharmonic of the qubit frequency. Arbitrary rotations on the Bloch sphere can then be realized with only $X/2$ gates combined with virtual Z-gates, which consist of a phase shift of subsequent gates [12] (Fig. 1B, bottom). This phase shift is accomplished by shifting the arrival time of these pulse trains with respect to the precession of the qubit state. While using SFQ pulses to control qubits has been previously demonstrated experimentally, the highest achieved Clifford gate fidelity utilizing locally generated SFQ signals was limited to the 95-98% range [13,14], and control of a multi-qubit system has yet to be demonstrated.

In this work, we demonstrate a novel active QPU, in which the quantum chip is integrated with a classical control chip based on SFQ digital technology via flip-chip bonding [15]. Fig.

1C shows the completed QPU in a multi-chip module (MCM) where the smaller quantum chip sits directly on top of the control chip. The spacing between the chips is precisely controlled to be 10 μm by the height of the bumps. The quantum chip consists of five transmon qubits, with a central fixed-frequency qubit connected to 4 outer tunable qubits via fixed capacitive coupling as shown in Fig. 1D. Previous experimental demonstrations of SFQ-based qubit control have consisted of converting an RF clock pulse to an SFQ pulse train that is coupled directly to a qubit [13,14,16]. Here, we add digital logic between the clock signal and the qubits in the form of a 1:4 demultiplexer (DMX). By utilizing time-domain multiplexing (TDM), we can use the same circuitry and clock line to control the outer 4 qubits. The center qubit is controlled by an individual SFQ driver (SD).

## SFQ-1Q Gate Results

We start by evaluating SFQ-based single-qubit (1Q) gates and discussing main limitations and strategies for improving fidelities. Results in this section are from the SFQ driver directly connected to the center qubit of one of our 5-qubit (5Q) devices. The circuit consists of a DC-SFQ converter that converts the rising edge of a clock signal to an SFQ pulse, and a passive-transmission-line (PTL) transmitter which delivers the SFQ pulse to the qubit (Fig. 1E). We calibrate 1Q gates (described in the Supplementary Material) and perform interleaved randomized benchmarking (IRB) [17] to determine average Clifford fidelity ($F_{Cliff}$) and $X/2$ fidelity ($F_{X/2}$). Averaged across 24 runs, we find $F_{Cliff}$ above 99.5%, and $F_{X/2}$ above 99.8%, a significant improvement from the previously best reported $F_{Cliff}$ of 98.4% [14]. These results, along with the RB curves and other interleaved fidelities from the highest fidelity run ($F_{X/2} > 99.9\%$) are shown in Fig. 2A.

To understand sources of error, we perform simulations of a $X/2$ gate using circuit parameters extracted from our measured device, while varying the coupling capacitance and the qubit coherence times (setting $T_1 = T_2$). The results of these simulations are shown in Fig. 2B (details given in Supplementary Material). At long gate times (small coupling capacitances), the fidelity is limited by decoherence, and at short gate times (large capacitances), it is limited by both leakage into the second excited state and discretization error. For a given coherence time, there is an optimal coupling capacitance (gate length) that minimizes the total error. At coherence times around the values measured across our current devices (20-80 μs), the optimal gate time is around 50 ns, which was the designed value and is comparable to the durations of conventional 1Q gates. Around these gate times, expected fidelities are around 99.8-99.9%, which agree well with our measured $F_{X/2}$ of 99.85(4)%. Fig. 2B also shows that the method of using resonant SFQ pulse trains (referred to as DANTE in [13]) requires coherence times above a millisecond in order to achieve fidelities above 99.99%, and this increase in performance is only achievable by increasing gate times significantly above those needed for state-of-the-art rf-based gates. It is therefore necessary to explore alternative pulse schemes that minimize leakage and shorten gate lengths to achieve fidelities on par with current state-of-the-art single-qubit performance, as shown in several theoretical works [18–22].

A primary concern with operating dissipative circuitry on a QPU is the generation of non-equilibrium quasiparticles [23], which can negatively impact qubit performance. In prior demonstrations, quasiparticle poisoning was shown to be the primary source of error, limiting gate fidelities to below 99% [13,14,24]. To investigate the impact of quasiparticle generation during operation, we clock the SFQ driver 30 MHz off-resonance immediately before performing a $T_1$ experiment (with the $\pi$ pulse coming from the SFQ driver). The resulting decay is fitted to both a standard exponential decay and a double exponential model that accounts for an average number of quasiparticles coupled to the qubit $\langle n_{qp} \rangle$ [25]. The results are shown in Fig. 2C. For poisoning times ($t_p$) up to 50 μs (~100,000 SFQ pulses at 1.8 GHz), there is no noticeable impact on $T_1$, and the extracted values of $\langle n_{qp} \rangle$ are quite low and difficult to accurately fit. Since coherence times and gate times are similar to those in prior works, these results suggest that the increase in our measured fidelity is a direct result of reduced quasiparticle poisoning. We attribute this absence of significant quasiparticle poisoning to several factors. Firstly, the chips are connected via bumps containing a large volume of aluminum which can act as a quasiparticle trap due to its lower superconducting gap, preventing quasiparticles from travelling to the qubit chip through the bumps [26]. Additionally, the qubit junctions used in this device are gap-engineered to mitigate quasiparticle tunneling across the junction barrier [27]. Finally, the PTL transmitter utilizes an underdamped junction to broaden the final SFQ pulse shape and reduce spectral content above frequencies that can break Cooper pairs directly in the qubit junction [28]. It is noted that as on-chip circuit complexity increases, power dissipation will also increase. Therefore, it will be required to continue to monitor and investigate quasiparticle dynamics and develop robust mitigation strategies for large-scale systems.

## Demultiplexer (DMX) Results

The outer 4 qubits on our device are controlled via a 1:4 DMX, made from a binary tree of dc-switches that connect the output of a single DC-SFQ converter to one of 4 PTL transmitters. The DMX channel is programmed with two dc current biases, connected via high-speed lines to allow for fast switching. Both programming biases have the same nominal value, with the selection bit determined by the sign of the bias.

Tune-up of the DMX consists of finding optimal bias values for the programming biases and the main circuit bias. Fig. 3A shows Rabi oscillations for each qubit/channel pair over a range of main bias currents, with each sweep showing a clear region of operation above some threshold bias and below the critical current of the circuit. The Rabi rates are fitted, and values falling within 5% of the median rate are shown in Fig. 3B. From this we can calculate a bias margin for the circuit of 6.2%, limited by the margin of ch0. To demonstrate dynamic switching, we perform the experiment illustrated in Fig. 3C, where we drive the 4 qubits in series before measuring the 4 qubits simultaneously. Results are shown in Fig. 3D and show the resulting Rabi oscillations, demonstrating that the DMX can be utilized within the context of a quantum algorithm, where channels must be changed at rates set by short gate durations. Some decay of the Rabi oscillations is seen in the qubits that idle for long periods between their drive pulse and readout. This suggests that gate compilation techniques and multiplexing schemes should be co-optimized so as to minimize idling errors during an intended algorithm.

From the Rabi rates we can quantify the crosstalk through the DMX, expressed in dB as $\Lambda_{ij} = 10\ \log_{10}\left(\frac{\Omega_{ij}}{\Omega_{jj}}\right)^2$, where $\Omega_{ij}$ is the Rabi rate on qubit $i$ when the DMX is set to channel $j$ [29]. The crosstalk matrix is shown in Fig. 3E. The average value of -35dB is close to the values reported in a similar MCM architecture [29]. Since the signal lines are completely shielded other than the exposed coupling pad, we attribute the crosstalk to some combination of stray capacitance to this pad and voltage crosstalk through the DMX. This crosstalk is not a limitation, unless qubits with nearly identical frequencies are connected to the same DMX.

## Gate Performance

Single-qubit gates through the DMX are calibrated using the same techniques as for the single-qubit driver. To characterize performance, we perform RB independently on each qubit (not simultaneously) utilizing a composition wherein each Clifford is composed of two $X/2$ gates, together with phase shifts to implement virtual-Z gates (referred to here as U3RB) [12]. With this composition, the $X/2$ error is half of the average Clifford error [30]. Fig. 4A shows the phase control on all 5 qubits by modulating the clock phase, with the 4 'lobes' as expected from driving at the second subharmonic. By measuring the final state with quantum state tomography (QST), we can also quantify the incoherent error of an average Clifford from the same sequence using Purity Randomized Benchmarking (PRB) [31], and traces for both U3RB and state purity for q0 are shown in Fig. 4B. The measured $X/2$ error for each qubit, along with the incoherent error is shown in Fig. 4C and reported in Table S1.

We find $X/2$ fidelities between 99.69-99.75%, with around half of the error coming from incoherent error. The measured incoherent error is similar in magnitude to the expected error from the measured $T_1$ and $T_2$ times. While a small portion of the remaining coherent error can be attributed to finite leakage ($< 1 \times 10^{-3}$, Fig. 2B), we attribute the bulk of the coherent error to under- and over-rotation errors due to an imprecise number of SFQ pulses being sent to the qubit. This imprecision originates from some variation in the number of clock edges that exceed the threshold current needed to trigger the DC-SFQ converter, which is a result of the limited sampling rate and bandwidth of the room-temperature arbitrary waveform generator (AWG) used to modulate the SFQ clock signal (1 GS/s and 300 MHz, respectively). To understand the impact of this imprecision, we simulate the fidelity of a variable number of pulses for various coherence times, shown in Fig. 4D. It is clear that a few missing or extra pulses can increase the error, and the impact becomes significant for qubits with longer coherence times (lower error floor). To eliminate this error mechanism and the need for amplitude modulation of the clock signal, we propose using a SFQ programmable pulse counter[32] to replace the external clock signal for triggering the DC-SFQ converter, shown in Fig. 4E. This 8-bit counter is self-resetting and serially programmable, with digital inputs that can be controlled with other SFQ circuits. Test waveforms from one of our 8-bit counters measured at 4K are shown in Fig. 4F, where pulse numbers are programmed into the counter, and corresponding pulse trains are generated when the circuit is triggered. We expect a significant improvement in fidelity from integrating this counter into our system.

To demonstrate the full functionality of our multi-qubit platform, we also perform conventional two-qubit (2Q) gates on our device by calibrating Controlled-Z gates based on Sudden-Net Zero flux pulses on the outer tunable qubits. We find fidelities between 97.4-99.5% using 2Q RB with all single-qubit gates performed via SFQ. Results are shown in the Supplementary Material.

## Power Dissipation

Power dissipation is a critical concern when placing active circuitry at the mixing chamber stage of the DR, since this is where the cooling power is at a minimum (typically ~20μW at 20mK), and excess heat will directly raise the temperature of the QPU. To evaluate the power dissipation of our technology, in Table 1, we compare the passive and active heat loads from our active SFQ circuitry with that of conventional "RF" wiring [5], and compare our SFQ DMX with a recently published "cryo-CMOS" DMX operating at mK [33].

Unlike CMOS circuits which require voltage biasing, SFQ logic cells require a current bias. The most common SFQ implementation, rapid SFQ (RSFQ), utilizes resistors to draw current from a common voltage bias [34]. While convenient, this bias scheme results in static power dissipation and presents a significant obstacle for operating at the millikelvin temperatures. In our system, we employe energy-efficient RSFQ (ERSFQ), which uses current-limiting Josephson junctions with series inductors rather than resistors, and can operate with zero static power dissipation when biased below the critical current of the current-limiting junction [35], the intended operating point for all our devices. The passive heat load is therefore determined entirely by the heat load of the associated wiring. For all cases, we only consider the passive heat load associated with the coaxial lines needed to clock the circuit, as this load is significantly greater than that of twisted pair dc lines typically used for dc biasing or digital inputs. For the "RF" case, we consider one coaxial line per qubit, and for the "cryo-CMOS" case, we use the reported static dissipation of 240 nW when operated at 0.7 V [33].

Since the DMX is programmed with dc-switches, there is no power dissipated when switching channels. Power is only dissipated during active operation (while the circuit is being clocked). We calculate the power consumed by the circuit ($P_D$) from the bias current ($I_b$) and the clock frequency ($f_{clk}$),

$$P_D = \Phi_0 I_b f_{clk}.$$

Assuming a clock frequency of 2.5 GHz ($f/2$ for a 5 GHz qubit), we find an active power dissipation of 9 nW for the 1:4 DMX, and 3 nW for the SD. Since the same DC-SFQ converter cell is used in both circuits, we can attribute 6 nW to the DMX circuitry itself (the dc-switches and additional PTL Transmitters). For the "RF" case, we simply use 4x the heat load of the single-qubit case, and for the "cryo-CMOS" case, the active dissipation comes from switching channels, which results in ~10 μW at 20 MHz (assuming switching between qubits every 50 ns, a typical gate length).

Additionally, we consider a 1:8 SFQ DMX, a simple extension of our existing design that needs only a single extra digital input line. This circuit requires a higher bias current and therefore an increased total heat load of 19 nW, but the load per qubit decreases ~30% from 3.25 nW/qubit to 2.375 nW/qubit. These values compare very favorably to the 5.5 nW/qubit in the “RF” case and are three orders of magnitude lower than the 2.5 μW/qubit in the “cryo-CMOS” case.

It is worth noting that the dc-switch based demultiplexer used in this work requires $\log_2 n$ fast current bias lines for $n$ DMX output channels. Using a fully digital demultiplexer based on non-destructive readout (NDRO) cells allows for digital programming of the output channel, replacing the fast current bias lines with digital input lines that can be controlled via additional digital circuitry [36]. This serial programming interface (also used in the programmable pulse counter) limits the required programming lines to only a serial data and serial clock line, regardless of the size of the demultiplexer. In addition, the clock rate needed to program an 8-bit counter (or 8-channel DMX) in 20 ns is only 400 MHz, meaning bulky coaxial cables are unnecessary for this application.

## Conclusion

The foundation of today’s microelectronic industry is built on the heavily integrated chip technology that enables system scaling, both in transistor count and system performance. Quantum computing, to be practical and scalable, will likely require a similar chip-based solution with qubits and other components integrated on the same chip or through chip stacking technologies. This work demonstrates progress towards this goal by developing an active QPU where SFQ control electronics and qubits are integrated into a single multi-chip module. We benchmarked the highest reported SFQ-based single-qubit gate fidelities to date ($F_{Cliff} > 99.5\%$) and implemented a digital demultiplexer to distribute gates to multiple qubits. Besides alleviating I/O scalability issues, the energy consumption of operating a large-scale QC will be significantly reduced by using a chip-based solution. Our DMX reduces the total heat load per qubit to below that of conventional RF wiring.

Additionally, by leveraging mature manufacturing flow and packaging techniques developed in the CMOS industry, such an integrated solution will also significantly reduce the cost of building large-scale systems. By integrating cryogenic SFQ control electronics directly with the quantum layer, this work presents a significant step towards realizing scalable quantum computing systems.

**Author contributions:** C.J. and S.J.H. conceived the project. Experiments were performed by J.B., with assistance by K.B., K.P., J.R., C.J., A.S. and A.W.. A.K. and C.J. designed the devices. C.J. and L.F.B. performed simulations. K.T.T., J.W., M.J.Y., M.R. and I.V. performed testing and validation of the carrier chip. J.R., K.B., L.F.B., K.P. and J.J. provided software support. C.C. fabricated quantum chips. I.V. and D.Y. managed testing and fabrication resources and facilities. S.J.H. oversaw the project. C.J. and S.J.H. wrote the manuscript with input from all authors.

**Competing interests:** The authors declare no competing interests.

**Data and materials availability:** All data, code, and materials used in the analysis are available upon reasonable request.

Supplementary Information is available for this paper.

Correspondence and requests for materials should be addressed to Shu-Jen Han.

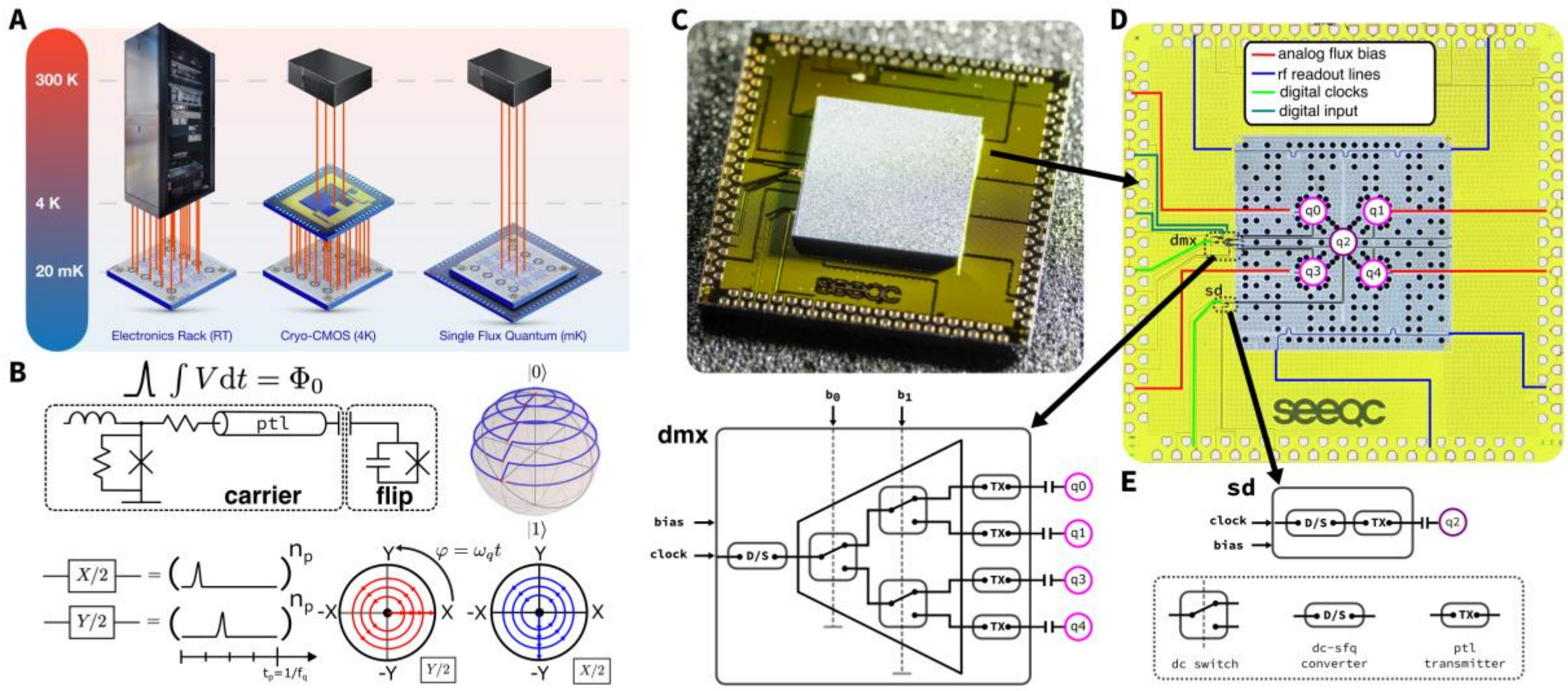

**Figure 1: System level overview. (A)** Comparison of conventional room-temperature control system (left), a conceptual 4K cryo-CMOS control system (middle), and an integrated SFQ control system (right). Only the integrated control system eliminates the linear scaling of control signals by implementing multiplexing at the final stage. **(B)** SFQ-based control of qubit state is realized by capacitively coupling SFQ voltage pulses to the qubit. Primitive $\pi/2$ gates consist of applying $n_p$ SFQ pulses at a rate commensurate with a qubit period. Phase control is realized by shifting the phase of the SFQ pulse trains with respect to the free precession of the qubit. For example, $X/2$ and $Y/2$ gates consist of the same pulse train, shifted by a quarter of a qubit period. **(C)** Photo of multi-chip-module consisting of a quantum flip-chip bump bonded to an SFQ carrier-chip. **(D)** Optical image of SFQ carrier-chip, with control and readout signal lines highlighted. **(E)** Component-level diagram of digital control circuits. The center qubit is controlled by a DC-SFQ converter which converts a clock edge into an SFQ pulse, which is delivered to the qubit via a passive-transmission-line transmitter. The four outer qubits are controlled with a 1:4 dc-switch based demultiplexer, with the output selected by 4 combinations of two current biases.

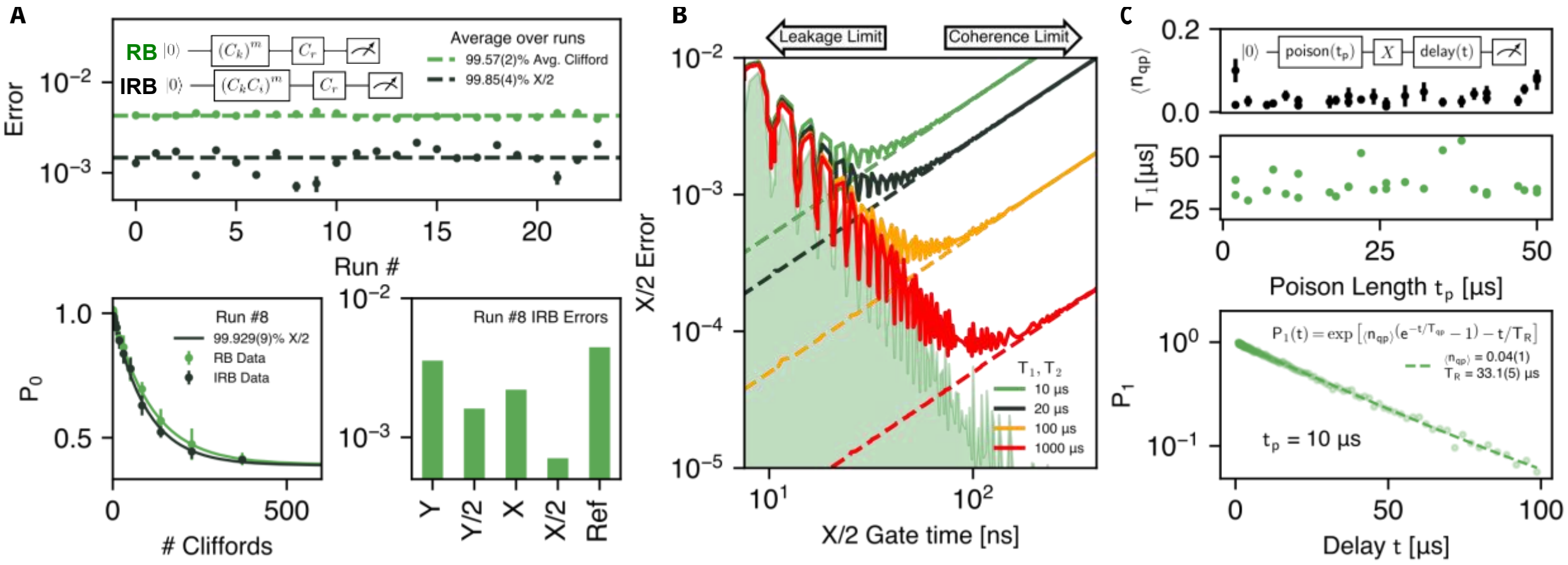

**Figure 2: Single-qubit gate performance. (A)** (top) Results from RB and IRB, showing average Clifford and $X/2$ fidelity over 24 runs, (bottom left) the full and interleaved RB traces from run #8 in the top plot, showing a fidelity above 99.9% for the $X/2$ gate, and (bottom right) the errors of other primitive gates from the same IRB run. Errors of $\pi$ rotations are larger than that of $\pi/2$ rotations due to their increased length. **(B)** Simulated error of a $X/2$ gate vs. gate time (controlled via coupling capacitance) for increasing coherence times. At long gate times (small capacitance), fidelity is limited by decoherence (theoretical coherence limit shown with dashed line), and at short gate times (large capacitance) fidelity is limited by leakage. **(C)** We generate a quasiparticle 'poisoning' pulse by clocking the SFQ circuit 30 MHz off-resonance for a duration $t_p$ and then perform a standard inversion recovery experiment. The resulting decay is fitted to a simple exponential to extract $T_1$, and to a double-exponential model (inset) to extract average quasiparticle number $\langle n_{qp} \rangle$. There is no obvious decrease in $T_1$ or increase in $\langle n_{qp} \rangle$ as poisoning length is increased up to 50 μs, and decay curves (bottom plot) do not show double-exponential behavior, suggesting quasiparticle poisoning is not increased above background levels.

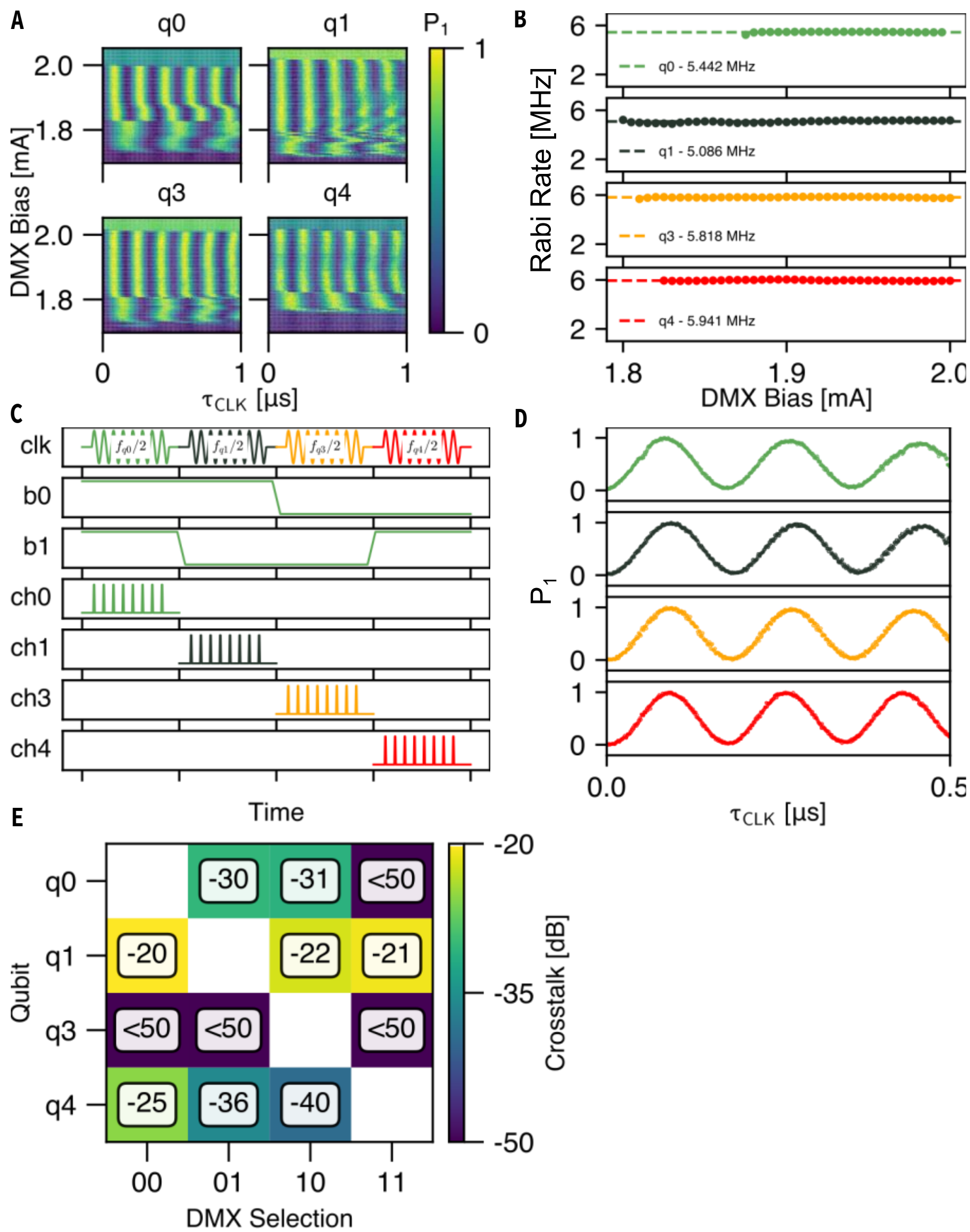

**Figure 3: Single-qubit control with SFQ demultiplexer. (A)** Rabi drive vs. DMX bias for all 4 qubits. **(B)** Extracted Rabi rate from the bias scans to determine operating margins. Points shown fall within a range of 5% around the median Rabi rate. **(C)** Timing diagram of time-multiplexed qubit control. To drive a qubit, the selection bits are set to the appropriate channel, and the clock is applied at a subharmonic of the target qubit frequency with the appropriate phase. Each clock edge generates an SFQ pulse, which is routed to the selected output channel. **(D)** Time-multiplexed Rabi drive of the 4 qubits. **(E)** Measured crosstalk through the DMX. Each qubit is clocked at its drive frequency for each DMX selection, with the resulting Rabi rate compared to the nominal rate through the correct channel.

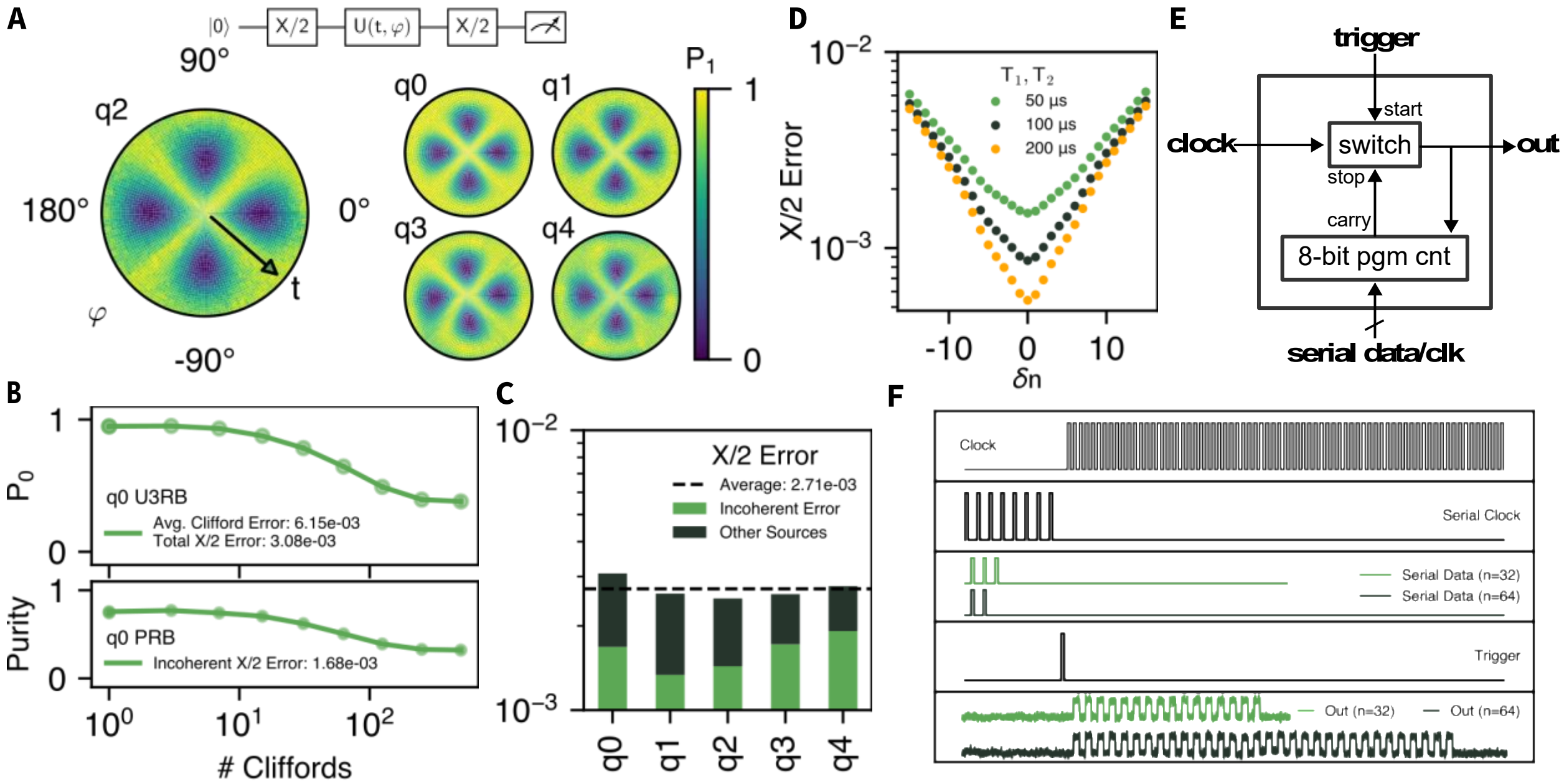

**Figure 4: Single-qubit gate performance through SFQ demultiplexer. (A)** Phase control of all 5 qubits, realized by modulating the phase of the SFQ clock signal. **(B)** U3RB and PRB traces from q0, used to quantify the total error, and the incoherent error associated with a $X/2$ gate, respectively. **(C)** Independently measured total and incoherent errors for all 5 qubits on the device. All qubits show similar error rates, with an average $X/2$ error of 2.7e-3. **(D)** Simulated error of an $X/2$ gate on q0 due to number of applied SFQ pulses, for increasing coherence times. Inaccuracy in pulse number results in over/under rotation that can significantly impact error at longer coherence times. **(E)** Block diagram for a pulse generator which can deliver a programmable number of SFQ pulses to the qubit. **(F)** Digital test results of a pulse generator measured at 4K, showing the programming and counting of 32 and 64 clock pulses (toggles on the output voltage).

| *Technology* <br> *# Qubits* | RF <br> 1 | SFQ <br> 1 | RF <br> 4 | Cryo-CMOS <br> 4 (DMX) | SFQ <br> 4 (DMX) | RF <br> 8 | SFQ <br> 8 (DMX) |
|---|---|---|---|---|---|---|---|
| *RF Lines* | 1 | 1 | 4 | 1 | 1 | 8 | 1 |
| *Digital IO Lines* | 0 | 0 | 0 | 2 | 2 | 0 | 3 |
| *DC Lines* | 0 | 1 | 0 | 1 | 1 | 0 | 1 |
| *Passive Heat Load (wiring) [nW]* | 4 | 4 | 16 | 4 | 4 | 32 | 4 |
| *Passive Heat Load (static dissipation) [nW]* | - | - | - | 240 | 0 | - | 0 |
| *Active Heat Load (2.5 GHz) [nW]* | 1.5 | 3 | 6 | - | 9 | 12 | 15 |
| *Switching Heat Load (20 MHz) [nW]* | - | - | - | $1 \times 10^4$ | 0 | - | 0 |
| *Total Heat Load [nW]* | 5.5 | 7 | 22 | $1.0244 \times 10^4$ | 13 | 44 | 19 |
| *Total Heat Load per qubit [nW]* | 5.5 | 7 | 5.5 | $2.56 \times 10^3$ | 3.25 | 5.50 | 2.375 |

**Table 1. Comparison of heat loads for SFQ vs. conventional and cryo-CMOS control technologies.** We compare the heat load of conventional ‘RF’ control, a recently published cryo-CMOS demultiplexer, and our SFQ-based demultiplexer for 1, 4, and 8 qubits. The heat load of the ‘RF’ scheme scales linearly with qubit number, while the SFQ-DMX scheme results in a decreasing heat load per qubit as the number of qubits increases, falling below the ‘RF’ case at just 4 qubits. Both of these cases are significantly lower than the reported results from the cryo-CMOS demultiplexer.

# Supplementary Information

## A Device Design and Fabrication

The devices reported are Multi-Chip Modules (MCMs) consisting of a 5mm x 5 mm chip containing quantum elements (the "quantum chip") bump-bonded to a 10mm x10mm chip containing all passive wiring and active SFQ elements (the "carrier chip"). The quantum chip consists of 5 transmon qubits and their readout resonators arranged in a star pattern with four outer qubits coupled to a common center qubit through a fixed capacitance. The four outer qubits are frequency-tunable, and the center qubit is fixed at a lower frequency.

The carrier chip contains two independent SFQ circuits for qubit control. A DC-SFQ converter is used to generate SFQ pulses to control the center qubit. For the four outer qubits, the output of a single DC-SFQ converter is routed through a 1:4 dc-switch based demultiplexer (DMX). A schematic of the DMX is shown and described in Fig. S3A. Both circuits are designed using ERSFQ technology resulting in zero static power dissipation. The SFQ pulses are delivered to the qubit via a passive transmission line (PTL), which has a small capacitance to the qubit. This circuit is described in Fig. S3B. Flux control of the qubits is realized via conventional high-speed flux-bias lines, and readout is performed using a common transmission line that is inductively coupled to the resonators on the quantum chip. All coupling of signals to the quantum chip is mediated through the vacuum gap between chips; there is no signal delivery through bump bonds. Other than elements which are required to directly couple to elements on the quantum chip, all wiring and circuitry on the carrier chip is covered with a superconducting "sky plane". Where openings in the sky plane must be made for coupling, the exposed dielectric underneath is etched so as to minimize exposure to the qubits and resonators.

The carrier chip is fabricated at Seeqc's foundry, using our SFQuClass process [37]. This process utilizes a 1kA/cm$^2$ Nb Josephson-junction process, a high-kinetic inductance layer, and additional Nb wiring and ground plane layers. Indium bump bonds are deposited on top of an Al hard-stop spacer which limits the inter-chip spacing to 10μm. The chips are bump-bonded together at room temperature [15], and wire-bonded into a PCB package for cryogenic testing.

## B Experimental Setup

Measurements were performed in a BlueFors LD dilution refrigerator with room temperature and cryogenic wiring shown in Fig. S1. The MCM was wire-bonded into a custom PCB package with both breakout SMP connectors for coaxial microwave signals and FFC connectors for low-speed signals. This package was mounted to the DR base plate inside of a custom magnetic shield, consisting of 2 concentric cylinders of μ-metal and 1 of copper (the cryo-can). At the top of the cryo-can, thermalized SMA feedthroughs and FFC-to-microD connectors pass the signals to higher stages. The input to the readout feedline was filtered by a K&L 8 GHz low-pass filter on the 4K stage, and an ECCOSORB filter within the μ-metal cryo-can and attenuated with 49dB of attenuation. The output of the readout feedline was amplified with a HEMT LNA on the 4K stage, preceded by isolators on the base plate. Since the die was designed with 2 readout feedlines to minimize trace crossings, these feedlines were daisy-chained by a coax within the cryo-can to create one readout path through the fridge. The SFQ circuits' clocks were provided by 29dB-attenuated RF lines, and their biases by DC lines, filtered by QDevil low-pass filters at the 4K stage. The DMX switches and the qubit flux-bias lines used 23dB-attenuated RF lines, with Mini-Circuits and K&L 8GHz low-pass filters mounted to the base plate outside the cryo-can, and ECCOSORB filters mounted inside the cryo-can.

SFQ clock signals were generated by heterodyne up-conversion of an LO provided by a SignalCore SC5510A, using a Marki IQ mixer and Keysight AWG for defining the frequency, phase, and length. The frequency-multiplexed readout tones were generated using the same up-conversion chain and acquired with a Keysight Digitizer after down-converting. DC current-biases for the SFQ circuits were applied with a Keysight Digital Stimulus and Response card (DSR). The experiment was orchestrated and triggered from a Keysight PXI chassis/embedded controller.

## C 1Q Gates

To implement single-qubit gates, we generate a train of SFQ voltage pulses which are capacitively coupled to the qubit [11]. Each voltage pulse rotates the qubit by a fixed amount $\delta\theta$. Using the perturbative model for the transmon as described in [38], we find an expression for $\delta\theta$ described by the capacitance to the qubit $C_C$, total qubit capacitance $C_q$, transmon zero-point fluctuation parameter $\xi = \sqrt{\frac{2E_C}{E_J}}$, and coefficient $\lambda \approx 1 - \frac{1}{8}\xi$.

$$\delta\theta = C_c\Phi_0\sqrt{\frac{2\omega_{10}}{\hbar C_q}} = 2\pi\frac{C_C}{C_q}\frac{\lambda}{\sqrt{\xi}}$$

Phase control is implemented by shifting the pulses in time, with respect to the qubit precession period. For example, a $Y_\pi$ gate is the same pulse train as an $X_\pi$ gate, delayed by a quarter of a qubit period. Example pulse trains are shown in Fig. 1B.

**Discretization Error**

The discrete nature of the SFQ pulses results in discretization error for arbitrary operations. However, the error of a gate is not linearly dependent on rotation error. For an arbitrary rotation angle $\theta$, a rotation error of $\Delta\theta$ results in gate error shown in Figure S4A. For example, for a $\pi/2$ pulse requiring $n_p = 100$ pulses ($\delta\theta = \pi/200$), the maximum discretization error is $\pi/400$, or 0.00125 radian, which results in gate error below 1e-5. Average gate error for a practical range of $n_p$ is shown in Fig. S4B.

**Loading Effects**

When a control line is coupled to a qubit, it introduces a loss channel (typically small and non-dominant), reducing the qubit $T_1$. This loading can be described as an additional loaded $Q_l$, quantified by the capacitance of the qubit mode $C_q$, the capacitance to the load impedance $C_c$, the impedance of the qubit $Z_q$, and the impedance of the load $Z_l$,

$$Q_l = \left(\frac{C_q}{C_c}\right)^2 \frac{Z_q}{Z_l}$$

As is shown in Fig. S3B, the matching resistors that prevent reflections in the PTL result in a voltage divider that divides the voltage pulse in half. Therefore we need a stronger coupling capacitance of $\sim 200\mathrm{aF}$ to make up for the reduced voltage. However, the load impedance is significantly lower than conventional lines, typically $Z_l = Z_{PTL}/2 = 3\Omega$ (the qubit sees two $6\Omega$ loads in parallel). The resulting Purcell Time ($Q_l/\omega_{01}$) is shown in Fig. S4C-D as a function of $C_c$ and $n_p$. For the coupling reported in our device, this results in a Purcell Limit of over 1ms, resulting in a negligible impact to the total $T_1$.

## D Gate Calibration

Unlike in analog qubit control where many parameters of a waveform can be modified to calibrate a desired gate, here only the total length (discrete number of SFQ pulses), frequency, and phase are available for calibration. After finding a trial $\pi/2$ gate length from a Rabi oscillation, ORBIT is used to calibrate each of these three parameters [39]. In an ORBIT experiment a gate parameter is swept while performing a Randomized Benchmarking experiment with fixed sequence-length $n$ . The $n$ is chosen such that the measured $p(0)$ is most sensitive to changes in that parameter (i.e. a maximum from a curve resulting from a subtraction between the pre-ORBIT RB curve and an RB curve corresponding to the target fidelity). As with RB, $k$ randomizations of this experiment are averaged. Gate parameters that result in high average $p(0)$ are taken to correspond to higher fidelity gates.

## E Quasiparticle Number Fitting

The quasiparticle numbers shown in Fig. 2C are obtained by fitting standard $T_1$ decay curves to a double-exponential model

$$\mathrm{P}(1) = e^{\langle n_{qp}\rangle\left(e^{-t/T_{qp}}-1\right)-t/T_R},$$

where $\langle n_{qp}\rangle$ is the average number of quasiparticles coupled to the qubit, $T_{qp}$ is the decay rate associated with a single quasiparticle, and $T_R$ captures the decay not associated with quasiparticles. Fits are only considered valid if fit residuals (reduced $\chi^2$) are lower for the double-exponential fit than for a single exponential fit.

## F Coherence Limit Calculation

We calculate the fidelity limit due to decoherence as

$$F_{lim} = \frac{1}{6}\left[3 + \exp\left(-\frac{\tau_g}{T_1}\right) + 2\exp\left(-\frac{2\tau_g}{T_2}\right)\right], \quad (1)$$

for a gate duration of time $\tau_g$ [40]. Since $\pi$ gates are twice as long as $\pi/2$ gates, they have a lower coherence limit.

## G Randomized Benchmarking

For the Interleaved RB results presented in Fig. 2, each individual run consists of 5000 shots of a standard RB experiment and an IRB experiment. Each standard RB experiment consists of 16 random Clifford sequences of lengths $2^n$, with n being integers from 0 to 10. At each Clifford length, the recovery from the 16 randomizations is averaged. The Interleaved RB experiment consists of the same experiment, but with the target gate interleaved between each Clifford. This entire sequence of experiments is performed for 24 runs over many hours and reported in Fig. 2. A minimal Clifford composition is used consisting of $I, \pm X/2, \pm Y/2, \pm X, \pm Y$ gates.

For the U3RB and PRB discussed in Figure 4, a"U3" composition is used such that all Cliffords consist of two physical $X/2$ gates interleaved with virtual-Z gates, from which we calculate the error due to each $X/2$ gate to be half of the average Clifford error.

## H Purity Benchmarking

Purity Benchmarking (PRB) is a variant of RB that captures only incoherent, non-unitary gate errors [31,42]. These errors are defined as trace-reducing errors; i.e. those that reduce $P = Tr(\rho^\dagger \rho)$ for accumulated (final) RB qubit state $\rho$. PRB is experimentally the same as RB, except the sequence of Cliffords is followed by quantum state tomography (QST) rather than a simple one-axis $\langle Z \rangle$ measurement, allowing $P$ to be determined.

The average Clifford incoherent error $\epsilon_{PRB}$ (a portion of $1 - F_{cliff}$) is calculated by

$$\epsilon_{PRB} = \frac{D-1}{D(1-\sqrt{\gamma})}; \quad \gamma = \exp\left(-\frac{1}{\tilde{n}}\right)$$

Where $\tilde{n}$ is the decay constant from an exponential fit of $P(n)$ with $n$ the number of Cliffords, and $D = 2^{N_q}$ is the Hilbert dimensionality of the system.

## I 2Q Calibration and Benchmarking

We calibrate conventional Controlled-Z (CZ) gates between the outer qubits and the central qubit, utilizing a Sudden Net-Zero (SNZ) pulse scheme [43]. Square flux waveforms are applied to the tunable qubits through analog bias lines. The amplitude and duration of these pulses are swept to find the 11-02 avoided level crossing, where the CZ gate is to be performed. After a flux amplitude and duration are chosen, a Controlled Phase experiment is performed to find the correct single-qubit phase corrections to add, and to quantify the leakage from the computational subspace (Fig. S2A). Once a pulse has been found with correct phase accumulation and minimal leakage, 2Q Randomized Benchmarking is performed to quantify the CZ error. Results from all 4 pairs are shown in Fig. S2B, and show CZ errors between 97-99%, limited primarily by leakage. We additionally verify our CZ results using quantum process tomography [44], with an example measured $\chi$-matrix shown in Fig. S2C. We calculate a process fidelity of 89.4 % between the measured $\chi_{meas}$ and the ideal $\chi_{CZ}$.

## J Gate Error Simulations

Here we describe the simulations used to produce the data in Fig. 2B and Fig. 4D. The system Hamiltonian is given by $H(t) = H_q + H_{int}(t)$, where $H_q = -E_J \cos\hat{\phi} + \frac{\hat{Q}}{2C_\Sigma}$ is the transmon Hamiltonian, and $H_{int}(t)$ describes the interaction between the qubit and a classical, time-dependent voltage pulse, given by $H_{int}(t) = \frac{C_c}{C_\Sigma}\hat{Q}V_{SFQ}(t)$, where $C_c$ is the coupling capacitance between the transmon and the SFQ pulse source, and $C_\Sigma = C_q + C_c$ is the total qubit capacitance.

Using the numerical Lindblad solver in QuTiP [45] (and including relaxation and dephasing), we generate a propagator $U_{SFQ}$ that represents the application of a single SFQ pulse and free precession with total duration $\tau_{CLK} = 1/f_{CLK}$. In these simulations, $f_{CLK} = f_q/2$. The time evolution of an arbitrary density matrix can then be described as repeated application of $U_{SFQ}$,

$$\rho(t + N_p\tau_{CLK}) = U_{SFQ}^{N_p}(\rho(t))$$

with $N_p$ the number of applied SFQ pulses.

The fidelity is given by

$$F_{avg} = \frac{1}{6}\left[\sum_k Tr\left(M_k^\dagger M_k\right) + \sum_k |Tr(M_k)|^2\right]$$

where $M_k = P_c U_0^\dagger G_k P_c$, $P_c = |0\rangle\langle 0| + |1\rangle\langle 1|$ is a projector onto the qubit subspace, and $\{G_k\}$ are the set of Kraus operators that represent $U_{sfq}(\rho) = \Sigma_k G_k \rho G_k^\dagger$ [46]. To calculate the average leakage and seepage rates between the computational and excited state subspaces, we use the definition given by [47]

$$L_1(U_{SFQ}) = U_{SFQ}\left(\frac{P_c}{N_c}\right)$$

$$L_2(U_{SFQ}) = 1 - U_{SFQ}\left(\frac{P_e}{N_c}\right)$$

where $P_e = |2\rangle\langle 2| + |3\rangle\langle 3|$ is the projector onto the excited state space, and where we have $N_c = 2$ states in both the computational and excited state spaces.

In Fig. 2B, we sweep the coupling capacitance $C_c$, finding the optimal $N_p$ and calculating $F_{avg}$ and $L_1$. In Fig. 4D, we sweep the number of applied pulses $N_p$, and calculate the error $1 - F_{avg}$.

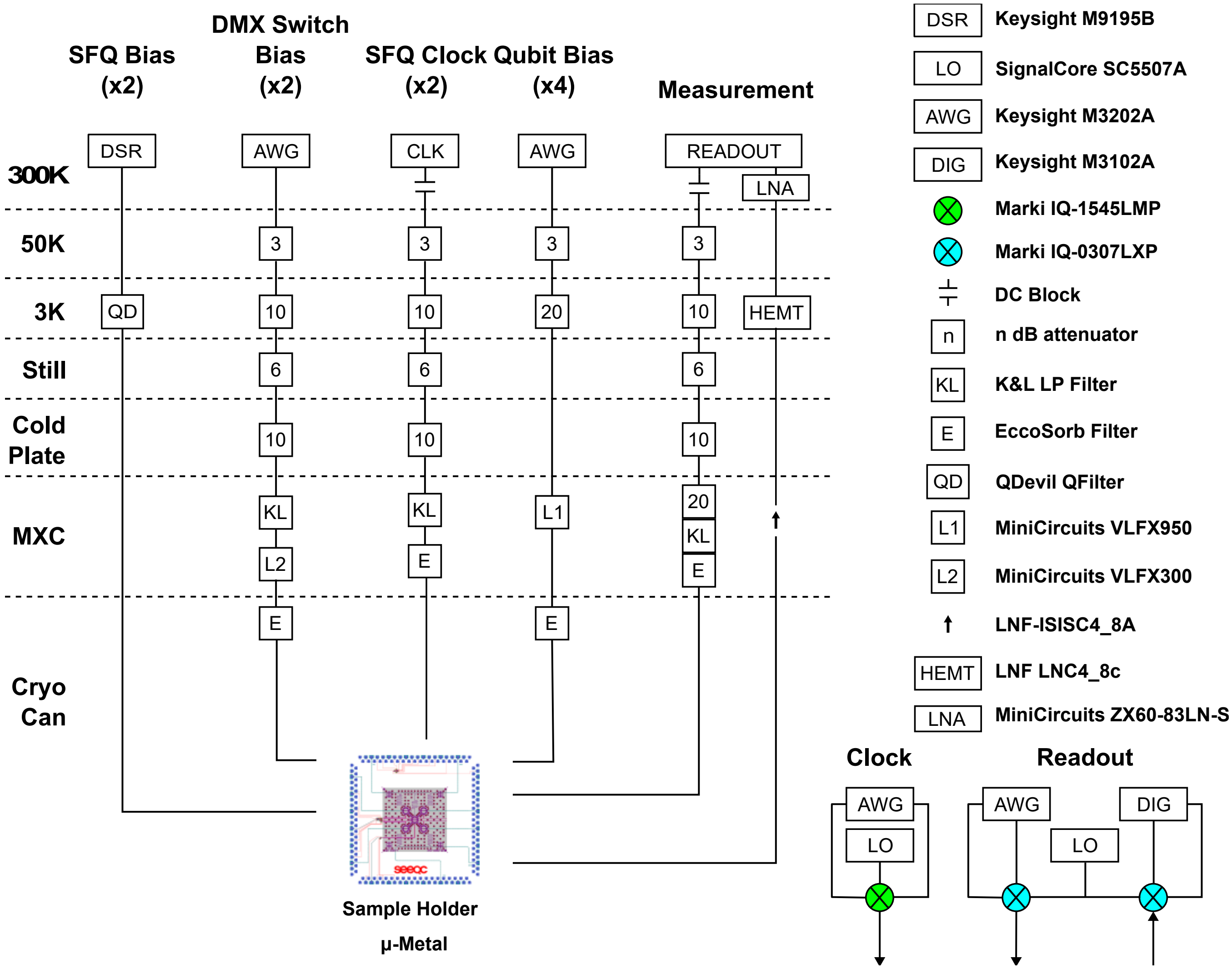

**Figure S1: Experimental Diagram**

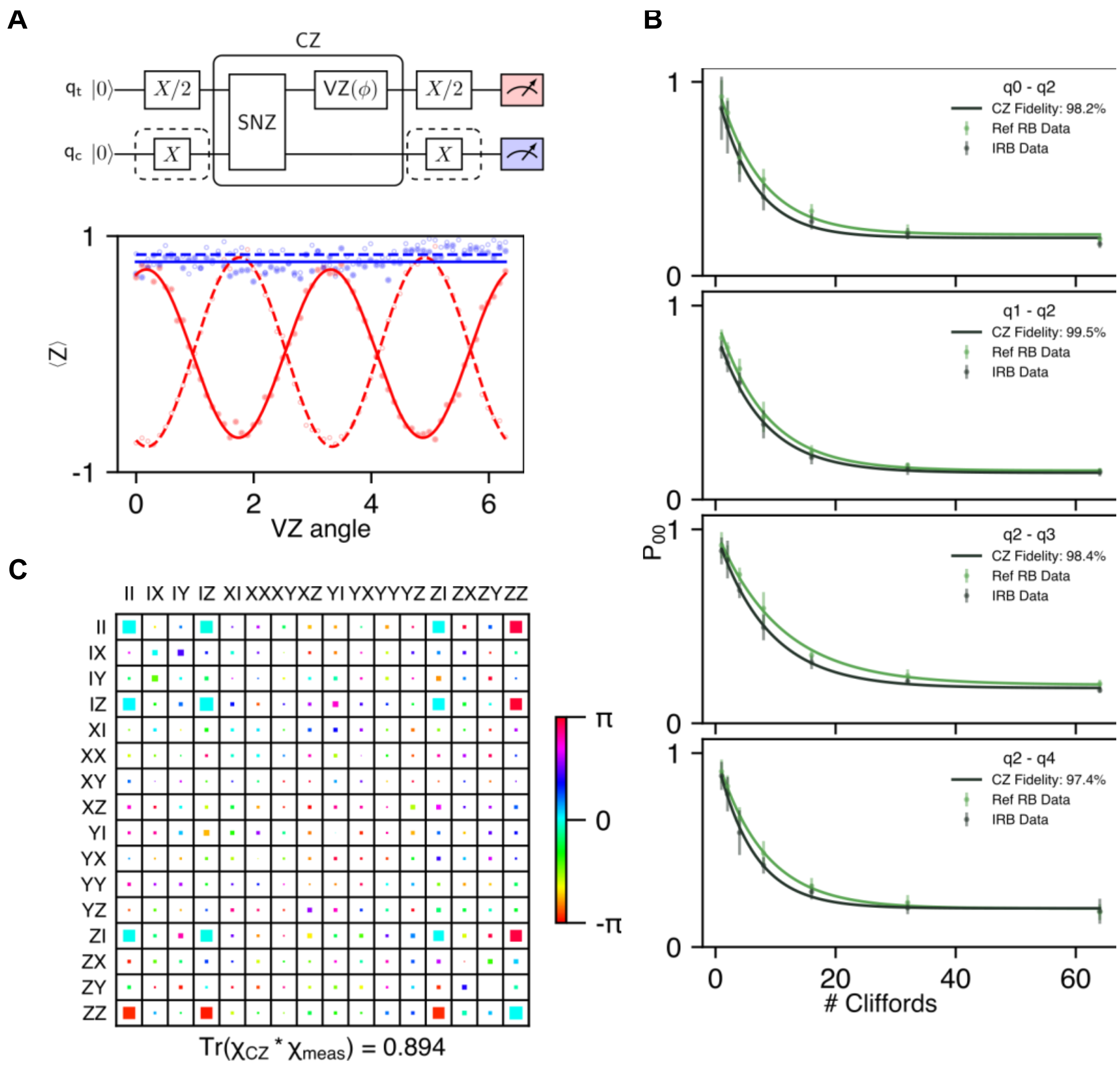

**Figure S2: 2Q Gate Results** (A) (top) Controlled Phase experiment between Q2 and Q3. A phase shift (Virtual Z gate) is added to the target qubit to correct accumulated phase. (bottom) Measurement of the target qubit (red traces) with (solid line) and without (dashed line) an $\boldsymbol{X}$ pulse applied to the control qubit, show a controlled phase difference of $\boldsymbol{\pi}$. Measurement of the control qubit (blue traces) with (solid line) and without (dashed line) an $\boldsymbol{X}$ pulse applied to the control qubit, showing leakage errors. (B) Interleaved 2Q Randomized Benchmarking on each pair, yielding CZ fidelities between 97.4-99.5%. (C) $\boldsymbol{\chi}$-matrix of the CZ gate between Q1 and Q2, with a process fidelity of 89.4%.

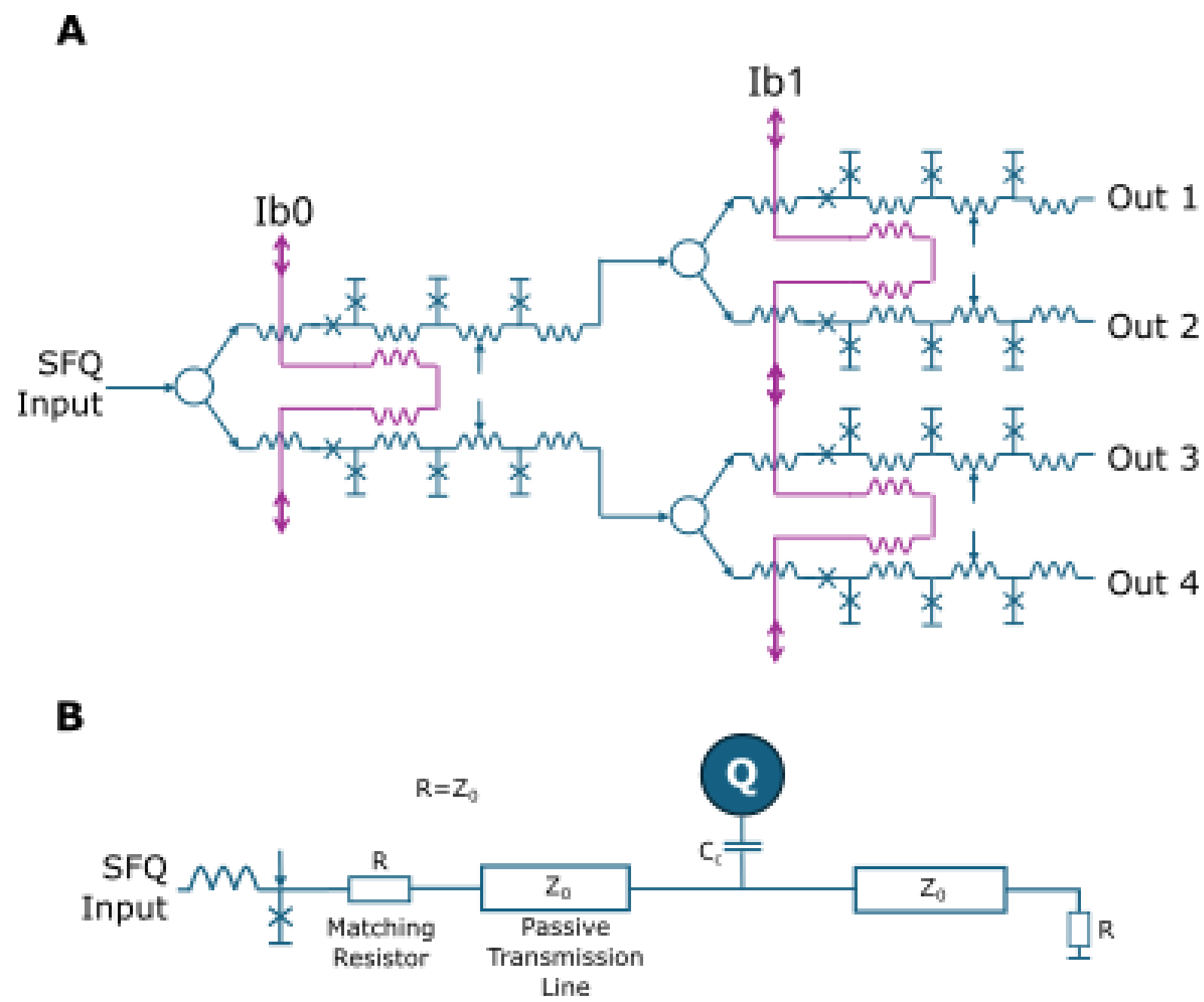

**Figure S3: SFQ Circuit Schematics** (A) Schematic of the DC-Switch based demultiplexer used in this work. The input signal is split between two dc-switches that are mutually coupled to a current bias line, such that a positive current opens only one channel and a negative current opens only the other. Each of these signal paths is split again through another layer of dc-switches, resulting in 4 channels selectable by the signs of the two bias currents $I_{b0}$ and $I_{b1}$. (B) Schematic of a PTL Transmitter (Tx) to deliver SFQ pulses to the qubit. The PTL Tx junction is underdamped such that the voltage pulse is broadened to reduce spectral content above 100 GHz. The pulse travels through a passive transmission line with a characteristic impedance of around $6\Omega$. The voltage pulse couples through a coupling capacitance $C_c$ to the qubit, before being dissipated in a matched resistor at the end of the line.

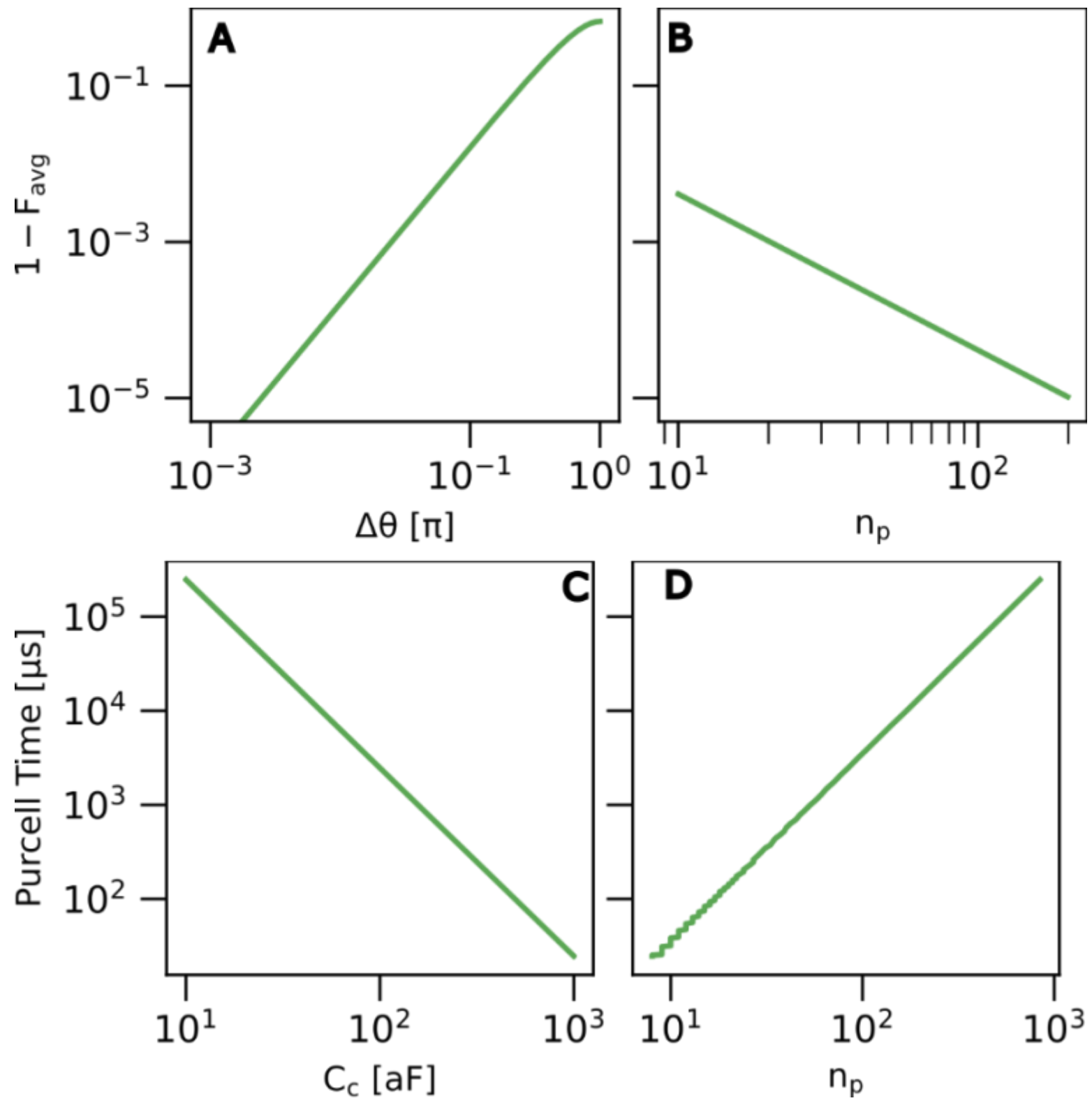

**Figure S4: Discretization error and loading effects** (A) Average gate error for a range of rotation errors such as that caused by discretization errors. (B) Average gate error due to the worst-case discretization error around practical values of $n_p$. For $n_p > 100$, discretization errors are well below 1e-4. (C) Purcell decay time from additional load into SFQ drive circuit for range in coupling capacitances. (D) Same data as (C), but plotted as function of $n_p$. For $n_p > 100$, the Purcell limit due to loss into the SFQ circuit is above 1ms.

| Parameter | Unit | Symbol | Method of determination | q0 | q1 | q2 | q3 | q4 |
|---|---|---|---|---|---|---|---|---|
| **Idle frequency** | GHz | $f_{01}$ | Ramsey spectroscopy | 5.083 | 4.794 | 3.800 | 4.684 | 4.826 |
| **Anharmonicity** | MHz | $\alpha_{01}$ | Two-tone spectroscopy | -280 | -287 | -291 | -285 | -277 |
| **Relaxation Time** | μs | $T_1$ | Inversion Recovery | 31 | 30 | 65 | 47 | 45 |
| **Decoherence Time** | μs | $T_{2E}$ | Hahn Echo | 12 | 13 | 26 | 10 | 22 |
| **SFQ Clock Rate** | GHz | $f_{clk}$ | $f_{01}/2$ | 2.542 | 2.397 | 1.900 | 2.342 | 2.413 |
| **Rabi Rate** | MHz | $\Omega_R$ | Fit to Rabi oscillations | 5.442 | 5.085 | 3.259 | 5.818 | 5.941 |
| **Number of SFQ pulses per $\pi/2$ gate** | | $n_{\pi/2}$ | $\frac{f_{clk}}{4\Omega_R}$ | 117 | 118 | 146 | 101 | 106 |
| **$\pi/2$ gate length** | ns | $\tau_{\pi/2}$ | 1Q Bringup | 46 | 49 | 77 | 43 | 44 |
| **Predicted fidelity limit due to decoherence** | % | $F_{Lim}$ | Eqn. (1) | 99.72% | 99.72% | 99.78% | 99.70% | 99.86% |
| **Measured coherence limit** | % | | Purity RB | 99.83% | 99.87% | 99.86% | 99.83% | 99.81% |
| **$X/2$ Fidelity** | % | $F_{X/2}$ | Randomized Benchmarking | 99.69% | 99.74% | 99.75% | 99.74% | 99.72% |
| **CZ gate length** | ns | $\tau_{CZ}$ | 2Q Bringup | 88 | 84 | - | 88 | 68 |
| **CZ Fidelity** | % | $F_{CZ}$ | 2Q Randomized Benchmarking | 98.2% | 99.5% | - | 98.4% | 97.4% |

**Table S1. Device properties.**